\documentclass[aps,prl,superscriptaddress,reprint,showpacs,longbibliography,amsmath,amssymb]{revtex4-1}
\usepackage[utf8]{inputenc}
\usepackage{microtype}
\usepackage{graphicx}
\usepackage{mathbbol}
\usepackage{amsfonts}
\usepackage{amssymb}
\usepackage{comment}
\usepackage[usenames,dvipsnames,svgnames,table]{xcolor}
\usepackage[colorlinks,linkcolor=blue,citecolor=blue,urlcolor=blue]{hyperref}

\def\equationautorefname~#1\null{Eq.~(#1)\null}

\usepackage{setspace}
\usepackage{braket}
\usepackage{physics} 
\usepackage{hyperref}
\usepackage[english]{babel}
\usepackage{color}
\usepackage{siunitx}

\newcommand{\bhat}{\hat{b}}
\newcommand{\bhatd}{\hat{b}^\dagger}

\newcommand{\od}{\omega_\mathrm{d}}
\newcommand{\om}{\omega_\mathrm{m}}
\newcommand{\ooc}{\omega_\mathrm{c}}
\newcommand{\ooL}{\omega_\mathrm{L}}

\newcommand{\kin}{\kappa_{\mathrm{in}}}
\newcommand{\kout}{\kappa_{\mathrm{out}}}

\newcommand{\kk}{\kappa}
\newcommand{\nc}{\bar{n}_\mathrm{c}}

\newcommand{\cd}{c_\mathrm{d}}
\newcommand{\cm}{c_\mathrm{m}}
\newcommand{\phid}{\phi_\mathrm{d}}
\newcommand{\phim}{\phi_\mathrm{m}}

\newcommand{\ahat}{\hat{a}}
\newcommand{\ahatd}{\hat{a}^\dagger}

\newcommand{\ii}{\mathrm{i}}

\begin{document}
	\title{Synthetic gauge fields for phonon transport in a nano-optomechanical system}
	\author{John P. Mathew}
	\affiliation{Center for Nanophotonics, AMOLF, Science Park 104, 1098 XG Amsterdam, The Netherlands}
	\author{Javier del Pino}
	\affiliation{Center for Nanophotonics, AMOLF, Science Park 104, 1098 XG Amsterdam, The Netherlands}
	\author{Ewold Verhagen}
	\email{verhagen@amolf.nl}
	\affiliation{Center for Nanophotonics, AMOLF, Science Park 104, 1098 XG Amsterdam, The Netherlands}
	\date{\today}

	\begin{abstract} 
	Gauge fields play important roles in condensed matter, explaining for example nonreciprocal and topological transport phenomena. Establishing gauge potentials for phonon transport in nanomechanical systems would bring quantum Hall physics to a new domain, which offers broad applications in sensing and signal processing, and is naturally associated with strong nonlinearities and thermodynamics.
In this work, we demonstrate a magnetic gauge field for nanomechanical vibrations in a scalable, on-chip optomechanical system.
We exploit multimode optomechanical interactions, which provide a useful resource for the necessary breaking of time-reversal symmetry. In a dynamically modulated nanophotonic system, we observe how radiation pressure forces mediate phonon transport between resonators of different frequencies, with a high rate and a characteristic nonreciprocal phase mimicking the Aharonov-Bohm effect. We show that the introduced scheme does not require high-quality cavities, such that it can be straightforwardly extended to explore topological acoustic phases in many-mode systems resilient to realistic disorder.
\end{abstract}

	\maketitle

Recent years have seen the emergence of theoretical and experimental efforts exploring exotic transport phenomena in bosonic systems exploiting broken structural and temporal symmetries~\cite{ozawa2018topological,goldman2016topological,Huber2016}.	
Magnetic gauge potentials play a particularly important role in those efforts. They can break time-reversal symmetry for transport, imparting a nonreciprocal, direction-dependent phase on a particle's wavefunction. For electrons, this leads to the celebrated Aharonov-Bohm effect~\cite{Aharonov1959} as well as the integer quantum Hall effect, which offers topologically protected transport in extended systems.
Periodic modulation has been proposed to create synthetic gauge fields to achieve similarly rich phenomena: It allows to effectively break time-reversal symmetry and explore emergent phases for electrons \cite{lindner2011floquet}, but importantly also for chargeless excitations of cold atoms and ions~\cite{bermudez2011synthetic,Goldman2014}, light~\cite{Fang2012b,Fang2012a} and mechanics~\cite{Nash2015}.
Cavity optomechanics provides a natural platform to realize time-varying potentials for either light or sound, and has been used to demonstrate nonreciprocal control of photons and phonons recently \cite{shen2016experimental,ruesink2016nonreciprocity,fang2017generalized,peterson2017demonstration,bernier2017nonreciprocal,xu2016topological,xu2018nonreciprocal}.
In many-mode optomechanical lattices, the phase and amplitude of interactions could be controlled with optical fields to induce exotic topological phases of light and sound at the nanoscale~\cite{Peano2015,Schmidt2015,Walter2016}. These inspiring proposals, however, require high quality factors and put extreme demands on fabrication tolerances and control intensities to achieve sufficiently strong interactions.

	\begin{figure}
		\centering
		\includegraphics[width=\columnwidth]{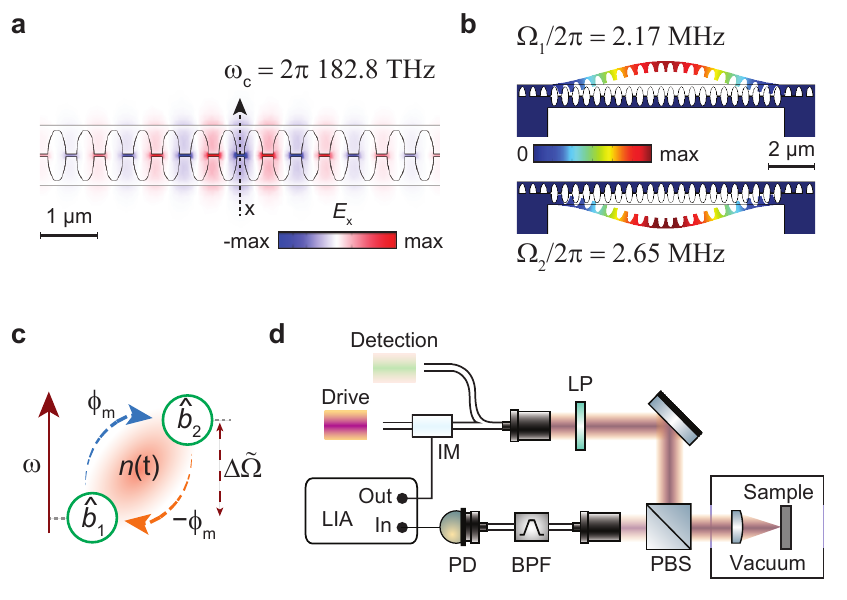}
		\caption{\textbf{Optomechanical system for synthetic nanomechanical gauge fields.} \textbf{a}, Simulated transverse electric field ($E_\mathrm{x}$) profile of the fundamental cavity mode of the sliced photonic crystal nanobeam. \textbf{b},  Displacement profile (exaggerated) of the mechanical modes employed in the experiment. 
			\textbf{c}, Time-modulated radiation pressure induces a synthetic gauge field for phonon transfer between modes of different frequency: The modulation phase is imprinted nonreciprocally along opposite transfer paths. Symbols as defined in text. \textbf{d}, Schematic of the experimental setup used. IM: intensity modulator, LP: linear polarizer, PBS: polarizing beam splitter, BPF: optical bandpass filter, PD: photodiode, LIA: lock-in amplifier. The LIA ports serve to (Out) drive the IM through an amplification stage (not shown) and to (In) analyze coherent intensity modulations of the detection laser.}
		\label{fig1}
	\end{figure}
	
	\begin{figure*}[t]
		\centering
		\includegraphics[width=2\columnwidth]{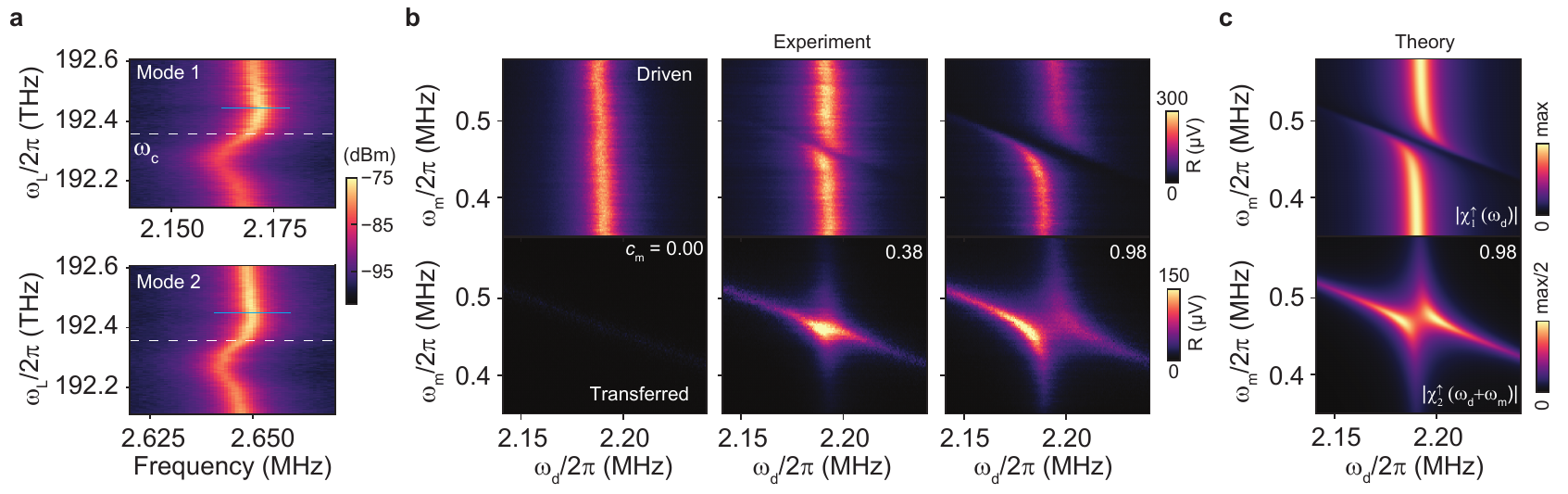}
		\caption{\textbf{Optically mediated phonon conversion.} \textbf{a}, Thermomechanical noise spectra of the two mechanical modes imprinted on the detection laser as the (unmodulated) drive laser frequency is swept across the cavity resonance. The mechanical modes are seen to be tuned via the optical spring shift. The white line denotes the cavity frequency $\ooc$ and the blue line denotes the drive laser frequency 
		used in subsequent measurements. Input power of the drive laser was locked to $P_\mathrm{in}=\SI{33.4}{\micro\watt}$ (see Methods). \textbf{b}, Amplitude of the driven response of mode 1 and simultaneously measured transferred response to mode 2 as a function of the modulation frequency for different modulation strengths $\cm$. Lock-in measurement of the detection laser intensity modulation at $\od\simeq\tilde{\Omega}_1$ gives the driven response and demodulation at $\od+\om\approx\tilde{\Omega}_2$ gives the transferred response. For $\cm=0$, there is no coherent transfer and only a faint thermal noise of mode 2 is observed. Transfer to mode 2 occurs in the vicinity of $\om \sim \Delta\tilde\Omega$. At large modulation strengths, a clear splitting is observed in the response. \textbf{c}, Theoretical response plots the absolute value of the susceptibility function for each mode for $\cm=0.98$ with phenomenological linewidths, $\Gamma_i/2\pi=\{8.1,8.6\}$ kHz (see Methods) where $\mathrm{max}=\mathrm{max}(\chi_1)$. The incident power in \textbf{b-c} is $P_\mathrm{in}=\SI{174}{\micro\watt}$.}
		\label{fig2}
	\end{figure*}
	
	Here we introduce a new mechanism to establish a magnetic gauge potential for sound at the nanoscale to overcome those challenges. It relies on optically-mediated mechanical mode transfer, which arises naturally in a system that dispersively couples two mechanical modes to a single optical cavity driven with a detuned laser~\cite{shkarin2014optically}: A displacement of one mechanical mode then shifts the cavity resonance, modifying the intra-cavity photon number and hence the radiation pressure acting on the second mode.
	This principle has been used to demonstrate coherent mechanical transfer and entanglement~\cite{weaver2017coherent,ockeloen2018stabilized}. 
Our experimental system is a sliced photonic crystal nanobeam~\cite{leijssen2015strong,leijssen2017nonlinear}, which supports an optical defect mode of frequency $\ooc=2\pi\times192.36$~THz and linewidth $\kk=2\pi\times256$~GHz that is localized in the subwavelength gap between the two halves of the nanobeam and coupled to the two $\sim2.4$~MHz-frequency in-plane flexural modes of the beam halves (cf. \autoref{fig1}\textbf{a}, \textbf{b}).

	To allow the creation of a gauge field, the lengths of the beam halves are kept dissimilar, which separates their mechanical resonance frequencies ($\Omega_i$, with $i=\{1,2\}$ labelling the mechanical modes) by about 0.5~MHz --- many times the mechanical linewidths $\Gamma_i\approx2\pi\times3$~kHz. 
Even though they are at very different frequencies, the two mechanical modes can be coupled through time-varying forces~\cite{okamoto2013coherent,faust2013coherent,weaver2017coherent,ockeloen2018stabilized,xu2018nonreciprocal}. By modulating the intensity of a drive laser incident from free space, the intracavity photon number acquires the form $n(t)=\bar{n}_\mathrm{c}(1+\cm\cos(\om t+\phim)+\cdots)$ with average photon population $\bar{n}_\mathrm{c}$, frequency $\om$, phase $\phim$ and modulation depth $\cm$, the dots denoting small modulation overtones at $(2k+1)\om,k\in\mathbb{N}$. Instantaneous response of photon number --- and thus the radiation pressure force --- to the incident modulated laser beam is guaranteed by the large cavity linewidth, meaning operation in the bad-cavity limit $\kappa\gg\Omega_i$.
	After linearization around the photonic steady state and adiabatic elimination of the cavity dynamics, the coherent evolution of the phononic modes (annihilation operators $\hat{b}_i$) within the resonance condition $\om=|\tilde{\Omega}_2-\tilde{\Omega}_1|$ is governed by the effective Hamiltonian in the mechanical rotating frame ($\tilde{b}_i=e^{\mathrm{i}\tilde{\Omega}_i t}\bhat_i$)
	\begin{equation}
	\tilde{H}_\mathrm{eff}=g_\mathrm{eff}(\tilde{b}^{\dagger}_1 \tilde{b}_2 e^{-\ii\phim}+\tilde{b}^{\dagger}_2 \tilde{b}_1 e^{\ii\phim})\label{eq:H_eff},
	\end{equation}
	where we set $\hbar=1$ (see derivation in Supplementary Information).

		In view of \autoref{eq:H_eff}, the modulated cavity field provides an effective pathway to swap phonons between mechanical modes $\hat{b}_1$ and $\hat{b}_2$, characterized by a nonreciprocal imprint of the modulation phase $\phi_\mathrm{m}$ for up/down mode conversion paths (see \autoref{fig1}\textbf{c}). As noted in photonics~\cite{Fang2012b,Fang2012a,Tzuang2014,Li2014a,roushan2017chiral} and cold atoms~\cite{goldman2016topological}, the modulation phase plays the role of a Peierls phase for a charged particle in a magnetic field. Here, the time-modulated radiation pressure force induces a synthetic magnetic flux for phononic transport, via an effective gauge potential defined through $\phim=\int_1^2\vec{A}_\mathrm{eff}\cdot\mathrm{d}\vec{l}$.

	In order to leverage this gauge field, it is paramount that the optically induced coupling rate $g_\mathrm{eff}$ overcomes mechanical dissipation as well as mechanical frequency disorder.
	The intermode coupling strength is given by $g_\mathrm{eff} = 2\cm g_1g_2\Delta/(\Delta^2+\kk^2/4)$, where $\Delta=\ooL-\ooc$ is the laser detuning and $g_i=g_{0i}\sqrt{\nc}$ denote the cavity-enhanced optomechanical couplings for photon-phonon coupling rates $g_{0i}$. We recognize here that the cross-mode coupling has similar origin as the optical spring effect, which shifts the mechanical frequencies to  $\tilde{\Omega}_i=\Omega_i+ 2g_i^2\Delta/(\Delta^2+\kk^2/4)$. 
	The coupling $g_\mathrm{eff}$ can overcome mechanical damping for typical detunings $\Delta\approx\pm\kappa/2$ if the cooperativity $4g_i^2/(\kk\Gamma_i)$ exceeds unity. This can be reached in the sliced nanobeam platform even for $\nc<1$~\cite{leijssen2017nonlinear}, and very large optical bandwidths~\cite{leijssen2015strong}. In fact, our unique advantage is the ability to operate
	in the bad-cavity limit $\kappa\gg\Omega_i$ --- in contrast to other proposed mechanisms~\cite{Peano2015,Schmidt2015,Walter2016} --- which greatly relaxes the practical requirements for extending the concepts to many-mode systems as shown below.

		In our experiment, we measure mechanical motion by analyzing intensity variations imprinted on a second `detection' laser far detuned from the cavity, using cross-polarized direct reflection (see \autoref{fig1}\textbf{d})~\cite{leijssen2015strong}. 
		The thermal fluctuation spectra of both mechanical modes are observed 
		in \autoref{fig2}\textbf{a}, showing the optical spring shift impacting both modes' frequencies $\tilde{\Omega}_i$ to similar extent as the drive laser is tuned across the cavity resonance. Subsequently, for a fixed detuning $\Delta=0.35\kk$, the drive laser intensity is modulated using the outputs of a lock-in amplifier (LIA) such that, besides the modulation tone, a weak probe tone $\cd\cos(\od t)$ is realized (with modulation depth $\cd\ll\cm$). Assuming without loss of generality that $\tilde{\Omega}_2>\tilde{\Omega}_1$ and tuning $\od\approx\tilde{\Omega}_1$, the mechanical mode 1 is driven and the strong tone at $\om\approx\Delta\tilde{\Omega}$ (defining $\Delta\tilde{\Omega}=\tilde{\Omega}_2-\tilde{\Omega}_1$) parametrically couples the two modes. Driven and transferred coherent responses of the mechanical modes are then analyzed by a lock-in measurement of the reflected detection laser (field $\alpha_\mathrm{d}(\od)$) intensity at $\od$ and $\od+\om$, respectively, providing information on the amplitude and phase response of the modes (see Methods). The outcome is captured by the linear susceptibilities $\langle {\hat{b}_{i}}(\omega)\rangle=\int\chi^{\uparrow\downarrow}_{i}(\omega-\omega')\alpha_\mathrm{d}(\omega')\hspace{0.5mm}\mathrm{d}\omega'$ ($\uparrow,\downarrow$ label  up- and down-transfer pathways), with 
		\begin{subequations}
		\begin{align}\label{eq:chis}
			\chi^\uparrow_1(\od)=&\frac{1+g_\mathrm{eff}\chi^{\uparrow}_{2}(\od+\omega_\mathrm{m})}{\od-\left(\tilde{\Omega}_{1}-\ii\frac{\Gamma_{1}}{2}\right)},\\
			\chi_2^\uparrow(\od+\om)=&\frac{g_\mathrm{eff}e^{i\phim}}{\left(\od-z_{+}\right)\left(\od-z_{-}\right)}.
		\end{align}
    	\end{subequations}
		Here, $z_\pm$ are the complex roots of the polynomial
		\begin{equation}\label{eq:zpm}
		\left[z-\left(\tilde{\Omega}_{1}-\ii\frac{\Gamma_{1}}{2}\right)\right]\left[z-\left(\tilde{\Omega}_{2}-\om-\ii\frac{\Gamma_{2}}{2}\right)\right]-g_\mathrm{eff}^{2}.
		\end{equation}
		
		 Without modulation ($\cm=0$), the coherent amplitude response of mode 1 is seen as a single peaked function in \autoref{fig2}\textbf{b}, with no corresponding transferred response. In contrast, as $\cm>0$, tuning the modulation frequency across $\Delta\tilde{\Omega}$ reveals broadening of the mode 1 response and non-zero transfer to mode 2 on resonance ($\om=\Delta\tilde{\Omega}$). The modes are hybridized, achieving strong coupling at the highest modulation strength ($\cm=0.98$), where a mode anticrossing is observed in the measured and predicted (\autoref{fig2}\textbf{c}) response amplitudes $|\chi_1^\uparrow(\od)|,|\chi_2^\uparrow(\od+\om)|$, with a resonant Rabi splitting of $\Omega^{\mathrm{eff}}_R=\sqrt{4 g_\mathrm{eff}^2-(\Gamma_1-\Gamma_2)^2/4}$. The strength of this dynamical strong coupling is tunable with the input optical power ($g_\mathrm{eff}\propto P_\mathrm{in}$) and coupling rates as high as 4\% of the mechanical frequencies are measured in experiment (see Supplementary Information). These large, tunable coupling rates, far exceeding achievable mechanical linewidths $\approx2\pi\times100$~Hz \cite{leijssen2017nonlinear}, emphasize the high potential of nanophotonic control to induce nontrivial forms of nanomechanical transport.

	\begin{figure}[t]
		\centering
		\includegraphics[width=\columnwidth]{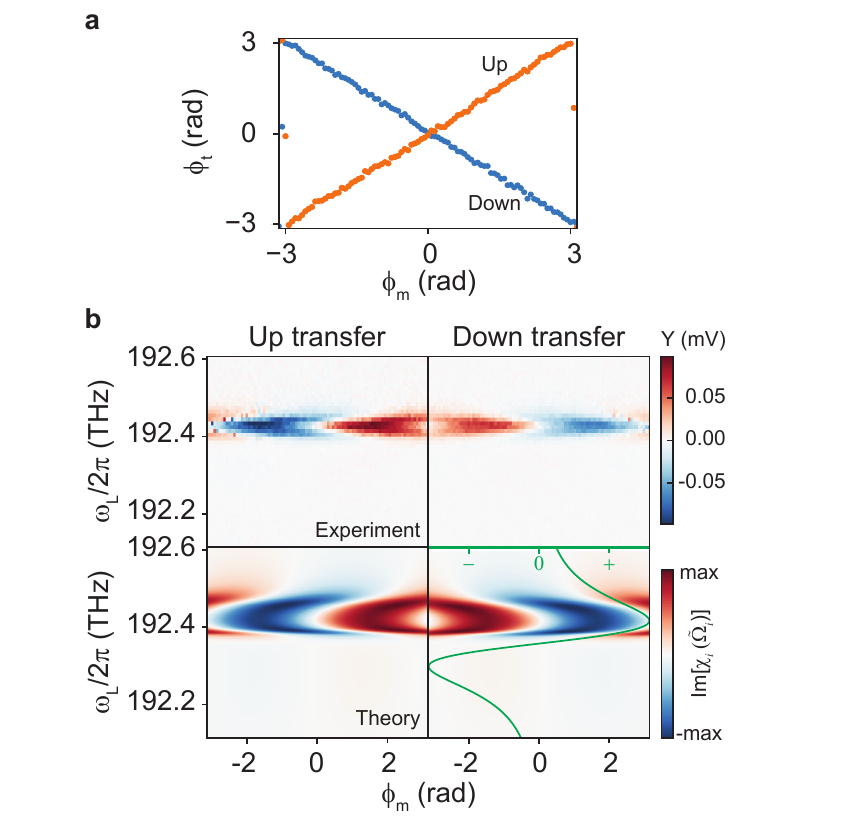}
		\caption{\textbf{Nonreciprocal phase imprint.} \textbf{a}, Phase measured at the frequency of the transferred mode for the up and down transfer processes as a function of the modulation phase, $\phim$.  
			The optically induced gauge field imprints a nonreciprocal modulation phase on the transfer. 
			\textbf{b}, The out-of-phase response of the two conversion paths as a function of the modulation phase and drive laser detuning. The theoretical response shows the imaginary part of the susceptibility of the transferred mode. The green curve and corresponding axis denote the calculated optical spring shift. For these measurements, the phase calibration procedure is followed at $\omega_\mathrm{L}/2\pi=\SI{192.45}{\tera\hertz}$ (see Methods). Modulation strength is $\cm=0.38$ in \textbf{a,b}.}
		\label{fig4}
	\end{figure}
	
	In order to experimentally address the phase acquired during nanomechanical transfer and the tunability of the gauge potential, the phase of the lock-in signal at the transfer frequency (for a fixed set $\od, \om, \cm$ tuned to achieve resonant transfer) is measured as a function of $\phim$ in Fig.~\ref{fig4}\textbf{a}. Since the input and detected phases evolve at different frequencies $\omega_\mathrm{d}$ and $\omega_\mathrm{d}\pm\omega_\mathrm{m}$, the transfer phases $\phi^{\uparrow\downarrow}$ cannot be unambiguously determined. They essentially depend on the (arbitrary) choice of origin of time $t$, which reflects gauge freedom. We effectively choose a certain gauge in an experimental calibration procedure (see Methods), by choosing the phase of the local oscillator to which the detected signal is referenced in the LIA such that the measured phase $\phi_\mathrm{t}=0$ when $\phi_\mathrm{m}=0$, before sweeping $\phi_\mathrm{m}$. In Fig.~\ref{fig4}\textbf{a} the transfer phase is seen to accumulate nonreciprocally during up- and down-conversion processes, as envisaged by \autoref{eq:H_eff}, confirming the role of $\phi_\mathrm{m}$ as a synthetic gauge potential for phonon transfer. These measurements are made possible by the fact that probe and modulation signals, as well as the local oscillator, are all referenced to the same clock in the LIA. As such, the mixing of local oscillator and detected signal at the target frequency can be interpreted as closing an Aharonov-Bohm loop composed of the nanomechanical transport (experiencing the gauge potential) and optical and electronic signal paths. 

	To further investigate the impact of the cavity mode in the process, the out-of-phase quadrature (see Methods) of the transferred response is measured as a function of laser frequency for both conversion pathways, and results are compared with the imaginary parts of the mechanical susceptibilities in \autoref{eq:chis}. 
	As observed in \autoref{fig4}\textbf{b}, phase nonreciprocity causes a change of sign in the imaginary part of the up/down transfer responses if $\omega_\mathrm{m}=\Delta\tilde{\Omega}$, i.e., $\Im\chi_{2}^{\uparrow}(\od^\uparrow+\om)=-\Im\chi_{1}^{\downarrow}(\od^\downarrow-\om)=g_\mathrm{eff}\sin\phi_\mathrm{m}/\left(g_{\mathrm{eff}}^{2}+\Gamma_{1}\Gamma_{2}/4\right)$ for $\phi^\uparrow=\phi_\mathrm{m}=-\phi^\downarrow$,  where $\od^\uparrow=\tilde{\Omega}_1$ and $\od^\downarrow=\tilde{\Omega}_2$. 
	Optimal transfer amplitude occurs at the maximum of the optical spring effect, with a non-zero transfer window for nearby frequencies determined by the spring shift of mechanical modes out of resonance for fixed $\od,\om$, yielding a non-zero phase $\varphi=\arg\left[(\tilde{\Omega}_1-z_{+})^{-1}(\tilde{\Omega}_2-z_{-})^{-1}\right]$ (see \autoref{eq:zpm}), accumulated for both up/down transfer channels and the break-down of the ideal gauge condition ($\phi^\uparrow\neq-\phi^\downarrow$).

\begin{figure}
	\centering
	\includegraphics[width=0.9\columnwidth]{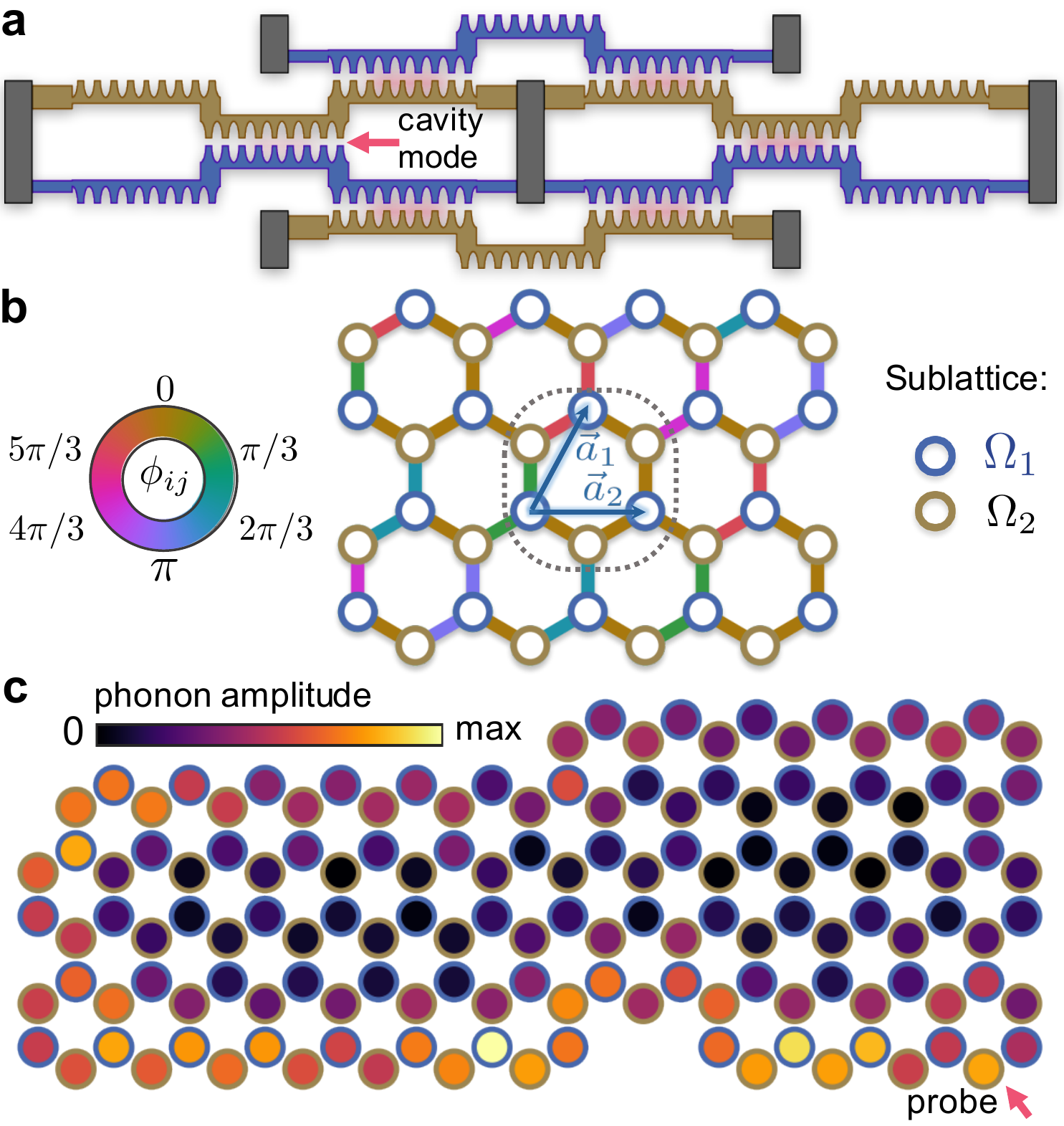}
	\caption{\textbf{Extended optomechanical nanobeam lattice.} \textbf{a}, Experimental realization of a nanobeam array. Nanobeams of different frequency are color-coded, with grey rectangles depicting supports. Each nanobeam is interconnected to three nanobeams of different frequency via broadband nanophotonic cavities. \textbf{b} This structure forms a honeycomb plaquette, surrounded by dashed lines, where each circle denotes an entire arm. By repetition, a honeycomb lattice with primitive lattice vectors $\vec{a}_1,\vec{a}_2$ is formed. The optical modulation phases to create a specific topological insulator phase, defined in the text, are indicated by the colors of the links. \textbf{c}, Simulated steady state phononic amplitude under continuous driving, focused on the site pointed by an arrow. Chosen parameters are $\Gamma_1/2\pi=\Gamma_2/2\pi=4.5$ kHz, $g_\mathrm{eff}/2\pi=200$ kHz, a phononic frequency disorder with standard deviation $\sigma_\Omega/2\pi=20$ kHz and a direct mechanical coupling of $t_2/2\pi=10$ kHz.}\label{Fig:4}
\end{figure}

  By virtue of bad-cavity limit operation of our optomechanical mechanism, the tolerances for nanobeam fabrication (see below), and the ability to optically control transport through out-of-plane illumination, the two-mode setup has high potential to be scaled up by suitably assembling many nanobeams. A synthetic gauge field for phonon transport could then enable artificial engineering of topological phases in extended systems which connect nanobeam resonators through THz-linewidth optical modes. As an example, we consider the 2D-network of nanobeams depicted in \autoref{Fig:4}\textbf{a}, where each mechanical mode is linked to three nearest neighbours of the opposite flavor (frequency) via modulated optical cavity modes, recognizing a honeycomb lattice with tunable links. By imprinting a spatially-varying phase pattern in the hopping terms, an artificial magnetic flux piercing the lattice can be realized. In particular, the phase choice $\phi_{ij}=2\pi\vec{r}_{ij}\cdot\vec{a}_2p/(aq)$ at lattice position $\vec{r}_{ij}$ ($a$ is the lattice constant and $p,q$ are coprime integers) along a unit vector $\vec{a}_2$ with no evolution along $\vec{a}_1$, (see \autoref{Fig:4}\textbf{b}) emulates the Landau gauge for an off-plane uniform magnetic field $B_\mathrm{eff}=2\pi/(qS)$, permeating each honeycomb plaquette with area $S$). 

  The 2D-extension of the phononic Hamiltonian \autoref{eq:H_eff} on resonance, namely
 \begin{equation}
 	 \tilde{H}^{\mathrm{lat}}_\mathrm{eff}=g_\mathrm{eff}\sum_{\langle i,j\rangle\in\mathrm{n.n.}}\tilde{b}^{(i)\dagger}_1 \tilde{b}^{(j)}_2 e^{-\ii\phi_{ij}}+\mathrm{H.c.},
 \end{equation}
 is dubbed as the Harper-Hofstadter model~\cite{Hofstadter1976}.
 Non-trivial topological properties are then signaled by 
 the formation of chiral edge states. For rational $p/q$ ($=1/3$ in the following), the bandstructure for an infinite ribbon geometry shows $2q$ bulk bands split into $2q-2$ gaps, traversed by counterpropagating edge states (see Supplementary Information). To exemplify the efficient propagation of phononic excitations along the boundary of the optomechanical array, we calculate the steady-state phononic amplitude for a finite lattice undergoing continuous-wave driving modulated at the frequency of a bandgap (see Supplementary Information for further details) and focused on a given site (\autoref{Fig:4}\textbf{c}). A moderate frequency disorder of $\sim 1\%$ is introduced, in addition to a direct mechanical coupling $\sim t_2\sum_{\langle i,j\rangle\in\mathrm{n.n.n.}}\sum_{l=1,2}(\tilde{b}^{(i)\dagger}_l \tilde{b}^{(j)}_l+\mathrm{H.c.})$ linking next-nearest neighbours ~\cite{leijssen2017nonlinear}. Disorder in photonic-mediated couplings is set to zero as it could be counterbalanced via intensity tuning of the incoming field (which could be derived from a single laser through judicious spatial structuring of amplitude and phase of its modulation sideband). As displayed by \autoref{Fig:4}\textbf{c}, even in the presence of imperfections, the topological protection of the edge state imprinted by the externally tailored gauge field results in a large unidirectional propagation length of phonons for parameters within state-of-the-art fabrication tolerances, demonstrating a feasible platform for on-chip phononic topological insulators.

In conclusion, we establish a synthetic gauge field for phonon transport using optically mediated couplings in an optomechanical system. We introduce an experimental platform with large optomechanical coupling strengths and bandwidths that leads to experimentally realizable many-mode implementations in the nanomechanical domain. The tunability of our system allows investigation of physics beyond the rotating wave approximation and in synthetic dimensions \cite{celi2014}. It opens up important avenues for exploring the impact of thermal and quantum fluctuations on topologically protected phonon transport~\cite{Rivas2017} and the effects of mechanical or optomechanical nonlinearities~\cite{leijssen2017nonlinear}. It is an advance towards exploiting topologically protected sound in the quantum acoustics regime and the search for exotic states such as those produced by non-Abelian gauge fields and analogs of fractional quantum Hall effect~\cite{Yuan2017} in the realm of nanomechanics. 


\section{Methods}
\subsection{Fabrication}
Devices were fabricated from a silicon-on-insulator substrate, with a 250 nm device layer and \SI{3}{\micro\meter} buried oxide layer (BOX). A 75 nm layer of diluted HSQ resist (1:2 in MIBK) was spin-coated, and electron-beam lithography was used to write patterns on the sample. After developing in TMAH, an anisotropic etch of the exposed device layer was done using ICP-RIE in the presence of HBr and O$_2$ gases. Finally, suspended nanobeams were obtained after a wet etch of the underlying BOX layer with hydrofluoric acid followed by critical point drying.

\subsection{Experimental setup}
The sample was placed in a vacuum chamber at room temperature and pumped down to a pressure of $\sim 2 \times 10^{-6}$~mbar. The devices were illuminated from outside the vacuum chamber using a broadband source and imaged in transmission on a phosphor coated NIR camera (imaging components not shown in \autoref{fig1}\textbf{d}). A tunable laser (Toptica CTL 1500) connected through a Thorlabs LN81S-FC intensity modulator (IM) was used as the drive laser, and a second laser (New Focus TLB-6728) far detuned from the cavity resonance ($\omega_\mathrm{L}/2\pi = 191.68$~THz) was used as the detection laser. The lasers were combined on a fiber-based beam combiner and launched using a fiber collimator into the free-space setup. Optical spring shifts were measured keeping the drive laser power locked by sampling the power in the free space path and feedback control on the bias port of the IM (power stabilization not shown in \autoref{fig1}\textbf{d}). Power stabilization was turned off for mode transfer measurements in order to operate the modulator at the optimal point (see Supplementary Information for modulation depth). The two outputs of the LIA carrying signals at $\omega_\mathrm{d}$ and $\omega_\mathrm{m}$ were combined, amplified (Mini Circuits ZHL-32A+ with 10 dB attenuation) and connected to the rf port of the IM to drive and modulate the nanobeam mechanics. The IM response ($V_\pi = 5.33$ V) and input amplification were characterized in order to quantify the modulation coefficients. The reflected detection laser was fiber coupled, filtered using a tunable bandpass filter (DiCon), and detected on a fast, low-noise photodetector. Intensity modulations of the detection laser were analyzed using a Zurich Instruments UHFLI lock-in amplifier (LIA).

\subsection{Lock-in measurement and phase calibration procedure}
The coherent mechanical transfer shown in the manuscript involves lock-in measurements performed by dual-phase demodulation of the detection laser signal at $\od\pm\om$, which provides the in-phase (X) and out-of-phase (Y) quadratures as referenced to the local oscillator of the LIA. For the phase measurements provided in \autoref{fig4}, a phase calibration procedure is  carried out to effectively define the time origin. The first step involves choosing the appropriate laser detuning and applying the probe and modulation tones, $V_\mathrm{d}=\cos(\od t + \phid)$ and $V_\mathrm{m}=\cos(\om t+\phim)$. Initially, the tones are applied from the two outputs of the LIA keeping $\phid=\phim=0$, with $\od=\tilde{\Omega}_1$ and $\om=\Delta\tilde{\Omega}$ (considering an up transfer measurement). The intensity modulations of the drive laser then give rise to phonon transfer that excites the second mode. This causes intensity modulations of the detection laser that are converted to an electronic signal (ignoring dc and $\tilde{\Omega}_1$ terms) of the form $V_\mathrm{sig}(t)=\sqrt{2} R\cos (\tilde{\Omega}_2t+\theta_\mathrm{arb})$ where $\theta_\mathrm{arb}$ arises due to timing differences in the applied tones. The action of the dual-phase, down-mixing performed by the LIA is mathematically represented by multiplication of the input signal with a complex local oscillator reference signal of the form $V_\mathrm{LO} (t)=\sqrt{2} \exp(\ii\omega_\mathrm{LO} t)$, where $\omega_\mathrm{LO}=\tilde{\Omega}_2$. The complex transfer signal after down-mixing and subsequent filtering is then given by $Z_\mathrm{sig}(t)=R\exp(\ii\theta_\mathrm{arb})$ which measures $\theta_\mathrm{arb}$. The local oscillator arm is now provided a phase shift equal to $\theta_\mathrm{arb}$ such that the new reference signal becomes $V_\mathrm{LO} (t)=\sqrt{2} \exp(\ii\omega_\mathrm{LO} t-\ii\theta_\mathrm{arb})$. The demodulated response of the transfer signal now has zero out-of-phase component and reads $Z_\mathrm{sig}(t)=R$. Subsequently, the phase of the modulation tone, $\phim$, is swept, causing the transferred signal to acquire the additional phase, $\phi_\mathrm{t}$, such that $V_\mathrm{sig}(t)= R\cos (\tilde{\Omega}_2t+\theta_\mathrm{arb}+\phi_\mathrm{t})$. The demodulated signal now becomes $Z_\mathrm{sig}(t)=R\exp(i\phi_\mathrm{t})$, which measures the transferred phase. The same procedure is then followed for the down transfer measurement.

\subsection{Mechanical linewidth broadening}
\noindent{} For the driven measurements presented here, the mechanical linewidths are observed to be larger than the intrinsic damping rates of $\sim2\pi\times3$~kHz. The subwavelength confinement of the optical mode and its colocalization with the mechanical modes causes sizable optomechanical cooperativities in our devices. In this scenario, the thermal motion of the mechanical modes causes large cavity frequency fluctuations of the order of $\kappa$, which in turn causes the optical spring shift to fluctuate. This leads to a broadening of the mechanical spectra beyond the intrinsic damping rate~\cite{leijssen2017nonlinear} and is captured in the model by using phenomenological linewidths $\sim2\pi\times8$~kHz. This spectral broadening can be overcome by increasing the optical linewidths or cooling the devices to reduce the effect of fluctuations.

\subsection{Simulation details for the edge state propagation}
 Continuous wave-driving focused on site $\vec{r}_0$ of a phononic lattice is introduced via  $\hat{H}_\mathrm{d}=\vec{\alpha}_\mathrm{d}(\omega_\mathrm{d},\vec{r}_0)\cdot\vec{\boldsymbol{\beta}}e^{-i\omega_\mathrm{L}t}+\mathrm{H.c.}$, where $w$ denotes the spot diameter and $\od$ stands for the frequency of the driving modulation tone, such that $\vec{\alpha}_\mathrm{d}(\omega_\mathrm{d},\vec{r}_0)\sim\cos(\omega_\mathrm{d} t) e^{-|\vec{r}_0|^2/w^2}$. Here $\vec{\boldsymbol{\beta}}$ is a vector encompassing phonon annihilation operators for localized phonons at different sites. The steady-state phononic amplitude ($\vec{u}_\mathrm{ss}$), in presence of phononic frequency disorder and next-nearest neighbour couplings obeys the sparse linear system, (writing $\hat{H}^{\mathrm{lat}}_\mathrm{eff}=\vec{\boldsymbol{\beta}}^{\dagger}\mathcal{A}\vec{\boldsymbol{\beta}}$),
 \begin{equation}
 \mathcal{A}\vec{u}_\mathrm{ss}=\vec{\alpha}_\mathrm{d}(\omega_\mathrm{d},\vec{r}_0),
 \end{equation}
  where the calculation of the matrix $\mathcal{A}$ is facilitated by use of the Kwant open source package~\cite{Groth2014}.

\bibliography{main_bib_abbv} 

\begin{thebibliography}{10}
\expandafter\ifx\csname url\endcsname\relax
  \def\url#1{\texttt{#1}}\fi
\expandafter\ifx\csname urlprefix\endcsname\relax\def\urlprefix{URL }\fi
\providecommand{\bibinfo}[2]{#2}
\providecommand{\eprint}[2][]{\url{#2}}

\bibitem{ozawa2018topological}
\bibinfo{author}{Ozawa, T.} \emph{et~al.}
\newblock \bibinfo{title}{Topological photonics}.
\newblock \emph{\bibinfo{journal}{Preprint at
  https://arxiv.org/abs/1802.04173}}  (\bibinfo{year}{2018}).

\bibitem{goldman2016topological}
\bibinfo{author}{Goldman, N.}, \bibinfo{author}{Budich, J.} \&
  \bibinfo{author}{Zoller, P.}
\newblock \bibinfo{title}{Topological quantum matter with ultracold gases in
  optical lattices}.
\newblock \emph{\bibinfo{journal}{{Nat. Phys.}}} \textbf{\bibinfo{volume}{12}},
  \bibinfo{pages}{639--645} (\bibinfo{year}{2016}).

\bibitem{Huber2016}
\bibinfo{author}{Huber, S.~D.}
\newblock \bibinfo{title}{{Topological mechanics}}.
\newblock \emph{\bibinfo{journal}{{Nat. Phys.}}} \textbf{\bibinfo{volume}{12}},
  \bibinfo{pages}{621--623} (\bibinfo{year}{2016}).

\bibitem{Aharonov1959}
\bibinfo{author}{Aharonov, Y.} \& \bibinfo{author}{Bohm, D.}
\newblock \bibinfo{title}{{Significance of electromagnetic potentials in the
  quantum theory}}.
\newblock \emph{\bibinfo{journal}{{Phys. Rev.}}}
  \textbf{\bibinfo{volume}{115}}, \bibinfo{pages}{485--491}
  (\bibinfo{year}{1959}).

\bibitem{lindner2011floquet}
\bibinfo{author}{Lindner, N.~H.}, \bibinfo{author}{Refael, G.} \&
  \bibinfo{author}{Galitski, V.}
\newblock \bibinfo{title}{Floquet topological insulator in semiconductor
  quantum wells}.
\newblock \emph{\bibinfo{journal}{{Nat. Phys.}}} \textbf{\bibinfo{volume}{7}},
  \bibinfo{pages}{490--495} (\bibinfo{year}{2011}).

\bibitem{bermudez2011synthetic}
\bibinfo{author}{Bermudez, A.}, \bibinfo{author}{Schaetz, T.} \&
  \bibinfo{author}{Porras, D.}
\newblock \bibinfo{title}{Synthetic gauge fields for vibrational excitations of
  trapped ions}.
\newblock \emph{\bibinfo{journal}{{Phys. Rev. Lett.}}}
  \textbf{\bibinfo{volume}{107}}, \bibinfo{pages}{150501}
  (\bibinfo{year}{2011}).

\bibitem{Goldman2014}
\bibinfo{author}{Goldman, N.} \& \bibinfo{author}{Dalibard, J.}
\newblock \bibinfo{title}{{Periodically driven quantum systems: Effective
  Hamiltonians and engineered gauge fields}}.
\newblock \emph{\bibinfo{journal}{{Phys. Rev. X}}}
  \textbf{\bibinfo{volume}{4}}, \bibinfo{pages}{1--29} (\bibinfo{year}{2014}).

\bibitem{Fang2012b}
\bibinfo{author}{Fang, K.}, \bibinfo{author}{Yu, Z.} \& \bibinfo{author}{Fan,
  S.}
\newblock \bibinfo{title}{{Photonic Aharonov-Bohm effect based on dynamic
  modulation}}.
\newblock \emph{\bibinfo{journal}{{Phys. Rev. Lett.}}}
  \textbf{\bibinfo{volume}{108}}, \bibinfo{pages}{153901}
  (\bibinfo{year}{2012}).

\bibitem{Fang2012a}
\bibinfo{author}{Fang, K.}, \bibinfo{author}{Yu, Z.} \& \bibinfo{author}{Fan,
  S.}
\newblock \bibinfo{title}{{Realizing effective magnetic field for photons by
  controlling the phase of dynamic modulation}}.
\newblock \emph{\bibinfo{journal}{{Nat. Photonics}}}
  \textbf{\bibinfo{volume}{6}}, \bibinfo{pages}{782--787}
  (\bibinfo{year}{2012}).

\bibitem{Nash2015}
\bibinfo{author}{Nash, L.~M.} \emph{et~al.}
\newblock \bibinfo{title}{Topological mechanics of gyroscopic metamaterials}.
\newblock \emph{\bibinfo{journal}{{Proc. Natl. Acad. Sci. USA}}}
  \textbf{\bibinfo{volume}{112}}, \bibinfo{pages}{14495--14500}
  (\bibinfo{year}{2015}).

\bibitem{shen2016experimental}
\bibinfo{author}{Shen, Z.} \emph{et~al.}
\newblock \bibinfo{title}{Experimental realization of optomechanically induced
  non-reciprocity}.
\newblock \emph{\bibinfo{journal}{{Nat. Photonics}}}
  \textbf{\bibinfo{volume}{10}}, \bibinfo{pages}{657--661}
  (\bibinfo{year}{2016}).

\bibitem{ruesink2016nonreciprocity}
\bibinfo{author}{Ruesink, F.}, \bibinfo{author}{Miri, M.-A.},
  \bibinfo{author}{Al\`{u}, A.} \& \bibinfo{author}{Verhagen, E.}
\newblock \bibinfo{title}{Nonreciprocity and magnetic-free isolation based on
  optomechanical interactions}.
\newblock \emph{\bibinfo{journal}{{Nat. Commun.}}}
  \textbf{\bibinfo{volume}{7}}, \bibinfo{pages}{13662} (\bibinfo{year}{2016}).

\bibitem{fang2017generalized}
\bibinfo{author}{Fang, K.} \emph{et~al.}
\newblock \bibinfo{title}{Generalized non-reciprocity in an optomechanical
  circuit via synthetic magnetism and reservoir engineering}.
\newblock \emph{\bibinfo{journal}{{Nat. Phys.}}} \textbf{\bibinfo{volume}{13}},
  \bibinfo{pages}{465--471} (\bibinfo{year}{2017}).

\bibitem{peterson2017demonstration}
\bibinfo{author}{Peterson, G.~A.} \emph{et~al.}
\newblock \bibinfo{title}{Demonstration of efficient nonreciprocity in a
  microwave optomechanical circuit}.
\newblock \emph{\bibinfo{journal}{{Phys. Rev. X}}}
  \textbf{\bibinfo{volume}{7}}, \bibinfo{pages}{031001} (\bibinfo{year}{2017}).

\bibitem{bernier2017nonreciprocal}
\bibinfo{author}{Bernier, N.~R.} \emph{et~al.}
\newblock \bibinfo{title}{Nonreciprocal reconfigurable microwave optomechanical
  circuit}.
\newblock \emph{\bibinfo{journal}{{Nat. Commun.}}}
  \textbf{\bibinfo{volume}{8}}, \bibinfo{pages}{604} (\bibinfo{year}{2017}).

\bibitem{xu2016topological}
\bibinfo{author}{Xu, H.}, \bibinfo{author}{Mason, D.}, \bibinfo{author}{Jiang,
  L.} \& \bibinfo{author}{Harris, J.}
\newblock \bibinfo{title}{Topological energy transfer in an optomechanical
  system with exceptional points}.
\newblock \emph{\bibinfo{journal}{Nature}} \textbf{\bibinfo{volume}{537}},
  \bibinfo{pages}{80--83} (\bibinfo{year}{2016}).

\bibitem{xu2018nonreciprocal}
\bibinfo{author}{Xu, H.}, \bibinfo{author}{Jiang, L.}, \bibinfo{author}{Clerk,
  A.} \& \bibinfo{author}{Harris, J.}
\newblock \bibinfo{title}{Nonreciprocal control and cooling of phonon modes in
  an optomechanical system}.
\newblock \emph{\bibinfo{journal}{Preprint at
  https://arxiv.org/abs/1807.03484}}  (\bibinfo{year}{2018}).

\bibitem{Peano2015}
\bibinfo{author}{Peano, V.}, \bibinfo{author}{Brendel, C.},
  \bibinfo{author}{Schmidt, M.} \& \bibinfo{author}{Marquardt, F.}
\newblock \bibinfo{title}{Topological phases of sound and light}.
\newblock \emph{\bibinfo{journal}{{Phys. Rev. X}}}
  \textbf{\bibinfo{volume}{5}}, \bibinfo{pages}{031011} (\bibinfo{year}{2015}).

\bibitem{Schmidt2015}
\bibinfo{author}{Schmidt, M.}, \bibinfo{author}{Kessler, S.},
  \bibinfo{author}{Peano, V.}, \bibinfo{author}{Painter, O.} \&
  \bibinfo{author}{Marquardt, F.}
\newblock \bibinfo{title}{Optomechanical creation of magnetic fields for
  photons on a lattice}.
\newblock \emph{\bibinfo{journal}{Optica}} \textbf{\bibinfo{volume}{2}},
  \bibinfo{pages}{635--641} (\bibinfo{year}{2015}).

\bibitem{Walter2016}
\bibinfo{author}{Walter, S.} \& \bibinfo{author}{Marquardt, F.}
\newblock \bibinfo{title}{{Classical dynamical gauge fields in optomechanics}}.
\newblock \emph{\bibinfo{journal}{{New J. Phys.}}}
  \textbf{\bibinfo{volume}{18}}, \bibinfo{pages}{113029}
  (\bibinfo{year}{2016}).

\bibitem{shkarin2014optically}
\bibinfo{author}{Shkarin, A.} \emph{et~al.}
\newblock \bibinfo{title}{Optically mediated hybridization between two
  mechanical modes}.
\newblock \emph{\bibinfo{journal}{{Phys. Rev. Lett.}}}
  \textbf{\bibinfo{volume}{112}}, \bibinfo{pages}{013602}
  (\bibinfo{year}{2014}).

\bibitem{weaver2017coherent}
\bibinfo{author}{Weaver, M.~J.} \emph{et~al.}
\newblock \bibinfo{title}{Coherent optomechanical state transfer between
  disparate mechanical resonators}.
\newblock \emph{\bibinfo{journal}{{Nat. Commun.}}}
  \textbf{\bibinfo{volume}{8}}, \bibinfo{pages}{824} (\bibinfo{year}{2017}).

\bibitem{ockeloen2018stabilized}
\bibinfo{author}{Ockeloen-Korppi, C.} \emph{et~al.}
\newblock \bibinfo{title}{Stabilized entanglement of massive mechanical
  oscillators}.
\newblock \emph{\bibinfo{journal}{Nature}} \textbf{\bibinfo{volume}{556}},
  \bibinfo{pages}{478--482} (\bibinfo{year}{2018}).

\bibitem{leijssen2015strong}
\bibinfo{author}{Leijssen, R.} \& \bibinfo{author}{Verhagen, E.}
\newblock \bibinfo{title}{Strong optomechanical interactions in a sliced
  photonic crystal nanobeam}.
\newblock \emph{\bibinfo{journal}{{Sci. Rep.}}} \textbf{\bibinfo{volume}{5}},
  \bibinfo{pages}{15974} (\bibinfo{year}{2015}).

\bibitem{leijssen2017nonlinear}
\bibinfo{author}{Leijssen, R.}, \bibinfo{author}{La~Gala, G.~R.},
  \bibinfo{author}{Freisem, L.}, \bibinfo{author}{Muhonen, J.~T.} \&
  \bibinfo{author}{Verhagen, E.}
\newblock \bibinfo{title}{Nonlinear cavity optomechanics with nanomechanical
  thermal fluctuations}.
\newblock \emph{\bibinfo{journal}{{Nat. Commun.}}}
  \textbf{\bibinfo{volume}{8}}, \bibinfo{pages}{16024} (\bibinfo{year}{2017}).

\bibitem{okamoto2013coherent}
\bibinfo{author}{Okamoto, H.} \emph{et~al.}
\newblock \bibinfo{title}{Coherent phonon manipulation in coupled mechanical
  resonators}.
\newblock \emph{\bibinfo{journal}{{Nat. Phys.}}} \textbf{\bibinfo{volume}{9}},
  \bibinfo{pages}{480} (\bibinfo{year}{2013}).

\bibitem{faust2013coherent}
\bibinfo{author}{Faust, T.}, \bibinfo{author}{Rieger, J.},
  \bibinfo{author}{Seitner, M.~J.}, \bibinfo{author}{Kotthaus, J.~P.} \&
  \bibinfo{author}{Weig, E.~M.}
\newblock \bibinfo{title}{Coherent control of a classical nanomechanical
  two-level system}.
\newblock \emph{\bibinfo{journal}{{Nat. Phys.}}} \textbf{\bibinfo{volume}{9}},
  \bibinfo{pages}{485} (\bibinfo{year}{2013}).

\bibitem{Tzuang2014}
\bibinfo{author}{Tzuang, L.~D.}, \bibinfo{author}{Fang, K.},
  \bibinfo{author}{Nussenzveig, P.}, \bibinfo{author}{Fan, S.} \&
  \bibinfo{author}{Lipson, M.}
\newblock \bibinfo{title}{{Non-reciprocal phase shift induced by an effective
  magnetic flux for light}}.
\newblock \emph{\bibinfo{journal}{{Nat. Photonics}}}  (\bibinfo{year}{2014}).

\bibitem{Li2014a}
\bibinfo{author}{Li, E.}, \bibinfo{author}{Eggleton, B.~J.},
  \bibinfo{author}{Fang, K.} \& \bibinfo{author}{Fan, S.}
\newblock \bibinfo{title}{{Photonic Aharonov-Bohm effect in photon-phonon
  interactions}}.
\newblock \emph{\bibinfo{journal}{{Nat. Commun.}}}  (\bibinfo{year}{2014}).

\bibitem{roushan2017chiral}
\bibinfo{author}{Roushan, P.} \emph{et~al.}
\newblock \bibinfo{title}{Chiral ground-state currents of interacting photons
  in a synthetic magnetic field}.
\newblock \emph{\bibinfo{journal}{{Nat. Phys.}}} \textbf{\bibinfo{volume}{13}},
  \bibinfo{pages}{146--151} (\bibinfo{year}{2017}).

\bibitem{Hofstadter1976}
\bibinfo{author}{Hofstadter, D.~R.}
\newblock \bibinfo{title}{{Energy levels and wave functions of Bloch electrons
  in rational and irrational magnetic fields}}.
\newblock \emph{\bibinfo{journal}{{Phys. Rev. B}}}
  \textbf{\bibinfo{volume}{14}}, \bibinfo{pages}{2239--2240}
  (\bibinfo{year}{1976}).

\bibitem{celi2014}
\bibinfo{author}{Celi, A.} \emph{et~al.}
\newblock \bibinfo{title}{Synthetic gauge fields in synthetic dimensions}.
\newblock \emph{\bibinfo{journal}{Phys. Rev. Lett.}}
  \textbf{\bibinfo{volume}{112}}, \bibinfo{pages}{043001}
  (\bibinfo{year}{2014}).

\bibitem{Rivas2017}
\bibinfo{author}{Rivas, {\'{A}}.} \& \bibinfo{author}{Martin-Delgado, M.~A.}
\newblock \bibinfo{title}{{Topological Heat Transport and Symmetry-Protected
  Boson Currents}}.
\newblock \emph{\bibinfo{journal}{{Sci. Rep.}}} \textbf{\bibinfo{volume}{7}},
  \bibinfo{pages}{1--9} (\bibinfo{year}{2017}).

\bibitem{Yuan2017}
\bibinfo{author}{Yuan, L.}, \bibinfo{author}{Xiao, M.}, \bibinfo{author}{Xu,
  S.} \& \bibinfo{author}{Fan, S.}
\newblock \bibinfo{title}{{Creating anyons from photons using a nonlinear
  resonator lattice subject to dynamic modulation}}.
\newblock \emph{\bibinfo{journal}{{Phys. Rev. A}}}
  \textbf{\bibinfo{volume}{96}}, \bibinfo{pages}{1--6} (\bibinfo{year}{2017}).

\bibitem{Groth2014}
\bibinfo{author}{Groth, C.~W.}, \bibinfo{author}{Wimmer, M.},
  \bibinfo{author}{Akhmerov, A.~R.} \& \bibinfo{author}{Waintal, X.}
\newblock \bibinfo{title}{{Kwant: A software package for quantum transport}}.
\newblock \emph{\bibinfo{journal}{{New J. Phys.}}}
  \textbf{\bibinfo{volume}{16}}, \bibinfo{pages}{063065}
  (\bibinfo{year}{2014}).

\end{thebibliography}


\begin{thebibliography}{8}%
		\makeatletter
		\providecommand \@ifxundefined [1]{%
			\@ifx{#1\undefined}
		}%
		\providecommand \@ifnum [1]{%
			\ifnum #1\expandafter \@firstoftwo
			\else \expandafter \@secondoftwo
			\fi
		}%
		\providecommand \@ifx [1]{%
			\ifx #1\expandafter \@firstoftwo
			\else \expandafter \@secondoftwo
			\fi
		}%
		\providecommand \natexlab [1]{#1}%
		\providecommand \enquote  [1]{``#1''}%
		\providecommand \bibnamefont  [1]{#1}%
		\providecommand \bibfnamefont [1]{#1}%
		\providecommand \citenamefont [1]{#1}%
		\providecommand \href@noop [0]{\@secondoftwo}%
		\providecommand \href [0]{\begingroup \@sanitize@url \@href}%
		\providecommand \@href[1]{\@@startlink{#1}\@@href}%
		\providecommand \@@href[1]{\endgroup#1\@@endlink}%
		\providecommand \@sanitize@url [0]{\catcode `\\12\catcode `\$12\catcode
			`\&12\catcode `\#12\catcode `\^12\catcode `\_12\catcode `\%12\relax}%
		\providecommand \@@startlink[1]{}%
		\providecommand \@@endlink[0]{}%
		\providecommand \url  [0]{\begingroup\@sanitize@url \@url }%
		\providecommand \@url [1]{\endgroup\@href {#1}{\urlprefix }}%
		\providecommand \urlprefix  [0]{URL }%
		\providecommand \Eprint [0]{\href }%
		\providecommand \doibase [0]{http://dx.doi.org/}%
		\providecommand \selectlanguage [0]{\@gobble}%
		\providecommand \bibinfo  [0]{\@secondoftwo}%
		\providecommand \bibfield  [0]{\@secondoftwo}%
		\providecommand \translation [1]{[#1]}%
		\providecommand \BibitemOpen [0]{}%
		\providecommand \bibitemStop [0]{}%
		\providecommand \bibitemNoStop [0]{.\EOS\space}%
		\providecommand \EOS [0]{\spacefactor3000\relax}%
		\providecommand \BibitemShut  [1]{\csname bibitem#1\endcsname}%
		\let\auto@bib@innerbib\@empty
		%
		\bibitem{Aspelmeyer2014}
        \bibinfo{author}{Aspelmeyer, M.}, \bibinfo{author}{Kippenberg, T.~J.} \&
          \bibinfo{author}{Marquardt, F.}
        \newblock \bibinfo{title}{{Cavity optomechanics}}.
        \newblock \emph{\bibinfo{journal}{{Rev. Mod. Phys.}}}
          \textbf{\bibinfo{volume}{86}}, \bibinfo{pages}{1391--1452}
          (\bibinfo{year}{2014}).
        %
		\bibitem{Reiter2012}
		\bibinfo{author}{Reiter, F.} \& \bibinfo{author}{S{\o}rensen, A.~S.}
		\newblock \bibinfo{title}{{Effective operator formalism for open quantum
				systems}}.
		\newblock \emph{\bibinfo{journal}{{Phys. Rev. A}}}
		\textbf{\bibinfo{volume}{85}}, \bibinfo{pages}{032111}
		(\bibinfo{year}{2012}).
		%
		\bibitem{SIleijssen2015strong}
		\bibinfo{author}{Leijssen, R.} \& \bibinfo{author}{Verhagen, E.}
		\newblock \bibinfo{title}{Strong optomechanical interactions in a sliced
			photonic crystal nanobeam}.
		\newblock \emph{\bibinfo{journal}{{Sci. Rep.}}} \textbf{\bibinfo{volume}{5}},
		\bibinfo{pages}{15974} (\bibinfo{year}{2015}).
		%
		\bibitem{abramowitz1964handbook}
		\bibinfo{author}{Abramowitz, M.} \& \bibinfo{author}{Stegun, I.}
		\newblock \bibinfo{title}{Handbook of mathematical functions: With formulas,
			graphs, and mathematical tables applied mathematics series}.
		\newblock \emph{\bibinfo{journal}{National Bureau of Standards, Washington,
				DC}}  (\bibinfo{year}{1964}).
	\end{thebibliography}
\bibliographystyle{NoURL_arXiv}

\noindent\\\textbf{Acknowledgements}\\
\noindent{}This work is part of the research programme of the Netherlands Organisation for Scientific Research (NWO). The authors acknowledge support from the Office of Naval Research (grant no. N00014-16-1-2466), the European Research Council (ERC Starting Grant no. 759644-TOPP), and the European Union's Horizon 2020 research and innovation programme under grant agreement no. 732894 (FET Proactive HOT).\\

\noindent\textbf{Author contributions}\\
\noindent{} J.P.M. performed the experiments and analyzed the data. J.d.P developed the theoretical model. E.V. supervised the project. All authors contributed to the interpretation of results and writing of the manuscript.\\

\clearpage
\pagebreak
\renewcommand{\theequation}{S\arabic{equation}}
\renewcommand{\thefigure}{S\arabic{figure}}
\onecolumngrid
\begin{center}
	\textbf{\large Supplementary Information}\\
	Synthetic gauge fields for phonon transport in a nano-optomechanical system
\end{center}

\setcounter{equation}{0}
\setcounter{figure}{0}
\setcounter{table}{0}
\setcounter{page}{1}
\makeatletter
\vspace{0.9cm}
\twocolumngrid

\section{I. Derivation of the effective Hamiltonian}
	The coherent dynamics of the system is modeled in the rotating frame of an input laser (detuned from the cavity by $\Delta$) by the optomechanical Hamiltonian for the two mechanical modes of interest $\hat{b}_i$, the cavity mode (annihilation operator $\hat{a})$, and their non-linear interaction measured by the vacuum coupling rates $g_{0i}$, reading
	\begin{multline}
		\hat{H}(t)=\sum\limits_i \Omega_i \bhatd_i\bhat_i- \ahatd\ahat\left[\Delta+\sum\limits_i g_{0i} (\bhat_i+\bhatd_i)\right]+\hat{H}_\mathrm{in}(t),\label{eq:H}
	\end{multline}
	where we set $\hbar=1$. The last term in \autoref{eq:H} introduces the time-varying input $\alpha_{\mathrm{in}}(t)$ with in-coupling rate $\kin$ and reads $\hat{H}_\mathrm{in}(t)=\ii\sqrt{\kin}\ahatd\alpha_\mathrm{in}(t) + \mathrm{H.c.}$. In the experiment, the intensity of the driving laser incident on the device is modulated using the high frequency outputs of the LIA (see \autoref{fig1}\textbf{d}) such that the relevant contribution is 
	\begin{equation}
	\frac{|\alpha^{\mathrm{eff}}_\mathrm{in}(t)|^2}{|\alpha_0^\mathrm{in}|^2}=1+\cm\cos(\om t+\phim)+\cd \cos(\od t),
	\end{equation}
	where $|\alpha_0^\mathrm{in}|^2$ is the average intensity of the incident field, and the modulation strength $\cm,\cd\in[0,2J_1(0.6\pi)\simeq1.163]$ ($J_{1}(z)$ is a Bessel function of the first kind) are the electrically controlled modulation depth factors at frequencies $\om$ and $\od$ respectively (further details below).
	In the strong field limit \autoref{eq:H} is linearized around the control input $\alpha_{\mathrm{in}}^\mathrm{m}(t)\propto c_\mathrm{m}\cos(\om t+\phim)$ by defining $\hat{a}\rightarrow\alpha^\mathrm{ss}_0(t)+\delta\hat{a}$ and neglecting small fluctuations $\mathcal{O}(\delta\hat{a}^2)$. Here $\alpha^\mathrm{ss}_0=\sqrt{\kin}\chi_{\alpha}(\omega)\alpha^{\mathrm{m}}_\mathrm{in}(\omega)$  is the photonic amplitude in the absence of the mechanical resonator or probe field ($g_{0i},c_\mathrm{d}=0$), and the optical susceptibility is defined as $\chi_{\alpha}^{-1}(\omega)=(\frac{\kappa}{2}-\ii(\omega+\Delta))^{-1}$ \cite{Aspelmeyer2014}. This yields $\hat{H}(t)\simeq\hat{H}_0(t)+\hat{V}^\mathrm{rp}(t)+\hat{V}^\mathrm{d}(t)$, where
	\begin{align}
		\hat{H}_0(t)	= & 	-\Delta \delta\hat{a}^{\dagger}\delta\hat{a}+\sum_{i}\big[\Omega_{i}\hat{b}_i^{\dagger}\hat{b}_i\hspace{-0.5mm}-\hspace{-0.5mm}|\alpha_{0}^{\mathrm{ss}}(t)|^{2}g_{0i}(\hat{b}_i+\hat{b}_i^{\dagger})\big],\\
		\hat{V}^{\mathrm{rp}}(t)	=&	\left(\alpha_{0}^{\mathrm{ss}}(t)\delta \hat{a}^{\dagger}+\mathrm{H.c.}\right)\left(\sum_{i}g_{0i}(\hat{b}_i+\hat{b}_i^{\dagger})\right),\label{eq:int1}\\
		\hat{V}^\mathrm{d}(t)	=&	\ii\sqrt{\kin}\left(\alpha_\mathrm{in}(t)\delta \hat{a}^{\dagger}+\alpha_\mathrm{in}(t)\alpha^{\mathrm{ss}*}_{0}(t)\right)+\mathrm{H.c.}.\label{eq:int2}\hspace{-1mm}
	\end{align}
	In the bad-cavity limit, intracavity population is instantaneously addressed by the external optical field
	\begin{equation}
	\alpha_{0}^{\mathrm{ss}}(t)\simeq\sqrt{\kin}\chi_{\alpha}(0)\alpha^{\mathrm{m}}_\mathrm{in}(t).
	\end{equation}
	For simplicity, we now consider the case where $\alpha_{\mathrm{in}}(t)$ only contains the strong term $\sim\cos(\om t+\phim)$. For moderate optomechanical coupling strengths and in the bad-cavity limit, the cavity dynamics reach the steady state rapidly and can be removed adiabatically. 
	
	The effective Hamiltonian thus follows from \cite{Reiter2012}
	\begin{equation}\label{eq:H_eff_general}
	H_\mathrm{eff}=\frac{1}{2}\hat{V}_{-}\sum_{f\in\{\pm\}}\left(\frac{\delta\hat{a}^{\dagger}\delta\hat{a}}{\Delta+\ii\frac{\kappa}{2}-\omega_f}+\mathrm{H.c.}\right)\hat{v}_{+}e^{f\om t} + \hat{H}_{g}.
	\end{equation}
	Here $\hat{V}_{\pm}$ (with $\hat{V}_+=\hat{V}_-^{\dagger}$) create/destroy excitations in the `excited' subspace (the high energy cavity modes in our case) and  $H_{g}=\sum_{i}\Omega_{i}b_{i}^{\dagger}\hat{b}_i-|\alpha_{0}^{\mathrm{ss}}(t)|^{2}(\sum_{i}g_{0i}(\hat{b}_i+\hat{b}_i^{\dagger}))$ governs the ground states evolution (mechanics). In \autoref{eq:H_eff_general}, the interactions in \autoref{eq:int1}, \autoref{eq:int2} are expanded into positive and negative frequency components, i.e. $	\hat{V}=\hat{v}_+(e^{\ii\om t}+e^{-\ii\om t})+\mathrm{H.c.}$. In the following we assume detuning is the largest frequency scale of the system ($\Delta\pm\om\simeq\Delta$), and the modulation frequency is assumed to be close to the frequency difference $\om\simeq\Delta\Omega$. Evaluation of \autoref{eq:H_eff_general} then yields the interaction Hamiltonian
	\begin{align}\label{eq:pre_RWA}
		\hat{H}_\mathrm{eff}^{\mathrm{int}}\simeq&\frac{\Delta}{(\Delta^{2}+\frac{\kappa^{2}}{4})}\{ |\alpha_{0}^{\mathrm{ss}}(t)|^{2}\left[\left(\sum_{i}g_{i}(\hat{b}_i+\hat{b}_i^{\dagger})\right)^{2}+\kin\right]+\nonumber\\&\ii\sqrt{\kin}\left[\alpha_\mathrm{in}^{*}(t)\alpha_{0}^{\mathrm{ss}}(t)+\mathrm{H.c.}\right]\sum_{i}g_{0i}(\hat{b}_i+\hat{b}_i^{\dagger})\}.\hspace{-0.8mm}	
	\end{align}
	Keeping the resonant terms in \autoref{eq:pre_RWA}, within the rotating-wave approximation (RWA) (in the picture defined as $\tilde{b}_i=e^{\ii\tilde\Omega_{i}t}\hat{b}_i)$ leads to
	\begin{equation}
	\label{eqnRotHam}
	\hat{H}_\mathrm{eff}(t) = \sum\limits_i \tilde{\Omega}_i  \bhatd_i  \bhat_i + g_\mathrm{eff}( \bhatd_1 \bhat_2 e^{-\ii\om t} e^{-\ii\phim}+ \mathrm{H.c.}),
	\end{equation}
	where the shifted frequencies and coupling rate, namely
	\begin{align}
		\tilde{\Omega}_i=\Omega_i+ \frac{2g_i^2\Delta}{\Delta^2+\kk^2/4},&&
		g_\mathrm{eff} = \frac{2g_1g_2\Delta}{\Delta^2+\kk^2/4}\cm,
	\end{align}
	contain the cavity-enhanced optomechanical coupling rates $g_i=g_{0i}\sqrt{\nc }$ with $\nc=\kin|\alpha^{\mathrm{in}}_{0}|^{2}/(\Delta^{2}+\frac{\kappa^{2}}{4})$ the 
	average intra-cavity population. The RWA, valid under the condition $g_\mathrm{eff}\ll\tilde{\Omega}_i$, effectively 
	removes terms creating and annihilating two excitations.
	
	\begin{figure*}[t]
		\centering
		\includegraphics[width=1\linewidth]{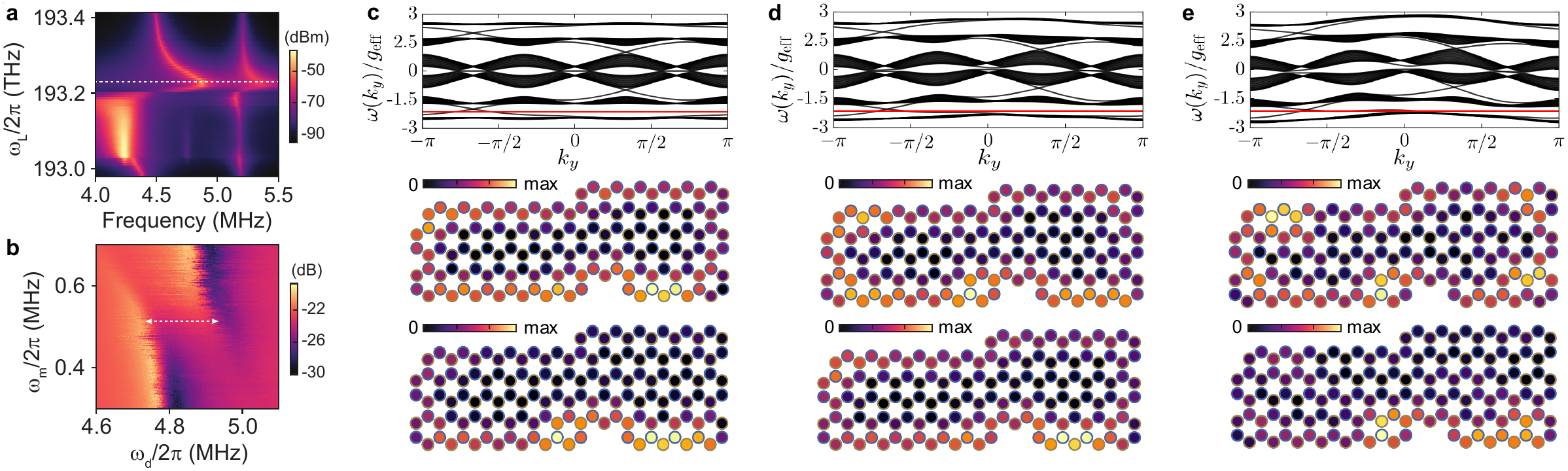}
		\caption{\textbf{Robust phononic edge states.} \textbf{a}, Optical spring shift measured at higher optical powers for a separate device shows large tunability of the mechanical modes. The dashed, white line shows the drive laser detuning used in \textbf{b}. The unusual features between 193.0 and 193.2 THz are due to a dynamically unstable regime of potential photothermoelastic origin.  \textbf{b}, The modulated coupling strength, $g_\mathrm{eff}$, is higher for higher optical powers. The largest modulated coupling strength measured in our experiments is marked by the dashed line and seen to be $\sim2\pi\times$200 kHz. Here the detection laser was absent and the drive laser response was directly demodulated using the LIA leading to a Fano shaped feature for the driven response. \textbf{c,d,e}, (top panels) Band structure for a ribbon geometry ($L=20$) for increasing values of the direct mechanical coupling from left to right ($t_2/2\pi=0,10$ and 20 kHz respectively), displaying the driving modulation frequency. The steady-state phononic amplitude in the absence of disorder is displayed in the middle row, while the result with a phononic frequency disorder with standard deviation $\sigma_\Omega/2\pi=20$ kHz (averaged over 100 realizations) is shown at the bottom. For this plots, $p/q=1/3$.}
		\label{SIfig3}
	\end{figure*}
	
	A similar treatment including modulation at a probe frequency (either $\tilde{\Omega}_1$ or $\tilde{\Omega}_2$) leads to an instantaneous term $\hat{H}_\mathrm{d}^{\mathrm{eff}}\propto \hat{b}_ie^{-\ii\od t}+\mathrm{H.c.}$. Cavity-induced losses in the mechanics (resulting in heating/cooling), arise in the formalism, but are neglected as $\Gamma_{i}\Delta/(\Delta^{2}+\frac{\kappa^{2}}{4})\ll\Gamma_{i}$.
	
	\section{II. Robustness of the phononic edge state transport}
	
	We here address in further detail the model for a 2D lattice of identical nanobeams. Assuming  \textit{i)} the bad-cavity limit holds for each of the optical links between nanobeam phononic modes (unit cells), \textit{ii)} the modulation frequency is tuned to the optimal transfer condition ($\om=\Delta\tilde{\Omega}$) and \textit{iii)} direct interaction between resonances of adjoining cavities is negligible, \autoref{eq:H_eff} generalizes, in the mechanical rotating frame, to
	\begin{align}
		\tilde{H}^{\mathrm{lat}}_\mathrm{eff}=	g_\mathrm{eff}^{(i)}\sum_{\langle i,j\rangle\in\mathrm{n.n.}}\tilde{b}^{(i)\dagger}_1 \tilde{b}^{(j)}_2 e^{-\ii\phi_{ij}}+\mathrm{H.c.}.\label{eq:HH_model}
	\end{align}
	The previous Hamiltonian reduces to the Harper-Hofstadter model for phase varying along a given direction (e.g. $\phi_{ij}=2\pi\vec{r}_{ij}\cdot\vec{a}_2p/(aq)$), a case that can be directly mapped to the Landau gauge field choice for a uniform magnetic field.
	
	Fabrication imperfections can influence the frequencies of the mechanical modes and introduce disorder into the system. This becomes relevant when considering optomechanical arrays for topological transport. From previous experiments on sliced nanobeams of same lengths, we estimate the frequency disorder to be less than 4\%, with possibility of further improvements using appropriate device designs and improved fabrication processes. Figure \ref{SIfig3}\textbf{a} shows parametric strong coupling achieved in a separate device where the Rabi frequency approaches \SI{200}{\kilo\hertz}. This value could be straightforwardly increased with increased optical powers. For the numerical simulations we considered a value $g_\mathrm{eff}=\SI{200}{kHz}$. Direct mechanical coupling between next-nearest-neighbour mechanical modes can be minimized by proper isolation of supports, as sketched in \autoref{Fig:4}\textbf{a}. It produces a contribution
	\begin{equation}
	\hat{H}^{(2)}_\mathrm{lat}= t_2\sum_{\langle i,j\rangle\in\mathrm{n.n.n.}}(\tilde{b}^{(i)\dagger}_1 \tilde{b}^{(j)}_1+\tilde{b}^{(i)\dagger}_2 \tilde{b}^{(j)}_2+\mathrm{H.c.}).
	\end{equation}
	
	For a ribbon geometry (size $\infty\times L$) the band structure of the Hamiltonian \autoref{eq:HH_model} is mirror-symmetric (owing to sublattice symmetry) and shows many gaps (the number, $2q-2$, is determined externally by the periodicity of the effective gauge potential along the direction $\vec{a}_2$). These are traversed by pairs of counter-propagating edge states (see the top panels in \autoref{Fig:4}\textbf{c,d,e}). As $t_2$ increases, energy gaps close and a localized excitation (delocalized in momentum space) hits bulk modes in addition to edge states (left to right panels). Phononic frequency disorder limits the average efficiency of the transport (bottom plots). Gap-closing events caused by mechanical coupling also imply that the system is more sensitive to disorder, as the transverse length of the edge states and scattering into the bulk is increased.
	
	\section{III. Experimental estimates and further details}
	\subsection{Electro-optic modulation and effective input driving}
	The input field intensity is modulated via an electro-optic intensity modulator, which produces an optical signal $|\alpha_\mathrm{in}(t)|^{2}=|\alpha_{0}^{\mathrm{in}}|^{2}(1+\Re[e^{i\pi(V/V_{\pi})}])$ where $V$ is an input electrical voltage. In our experiment, this contains three contributions: \textit{i)} a carrier tone at zero-frequency $V_\pi$, \textit{ii)} a modulated signal $V_\mathrm{m}$ and a \textit{iii)} weak probe voltage $V_\mathrm{d}$. Then $V=V_{\pi}(\frac{1}{2}+\frac{V_\mathrm{m}}{V_{\pi}}\cos\omega_\mathrm{m}t+\frac{V_\mathrm{d}}{V_{\pi}}\cos\omega_\mathrm{d}t$) and 
	\begin{align}
		\Re e^{i\frac{\pi V}{V_{\pi}}}\simeq&
		-\pi\frac{V_\mathrm{d}}{V_{\pi}}\cos\omega_\mathrm{d}t\cos\left(\pi\frac{V_\mathrm{m}}{V_{\pi}}\cos\om t\right)-\nonumber\\
		&\sin\left(\pi\frac{V_\mathrm{m}}{V_{\pi}}\cos\om t\right),
	\end{align}
	where $V_\mathrm{d}\ll V_{\pi}$ is assumed. Jacobi-Anger expansions \cite{abramowitz1964handbook} show the former contribution only includes even overtones $\sim\cos(2k\om t)$ while the latter contains only the odd modulation harmonics $\sim\cos((2k-1)\om t)$ ($k\in\mathbb{N}$). For modulation frequencies approaching resonance $\omega_\mathrm{m}\simeq\Delta\tilde{\Omega}$, the  terms that couple resonantly to the mechanics are conveniently arranged into
	$|\alpha_\mathrm{in}^\mathrm{eff}(t)|^{2}=|\alpha_{0}^\mathrm{in}|^{2}(1+\cm\cos\omega_\mathrm{m}t+\cd\cos\omega_\mathrm{d}t)$, where the modulation depth $\cm=2J_{1}(\pi\frac{V_\mathrm{m}}{V_{\pi}}),\cd=-\pi\frac{V_\mathrm{d}}{V_{\pi}}J_{0}(\pi\frac{V_\mathrm{m}}{V_{\pi}})$ measure the strength of the control and (weak) probe components. A non-zero phase $\phi_\mathrm{m}$ imprinted in the modulation, leaves the previous discussion unchanged, except for the replacement $\cos\omega_\mathrm{m}t\rightarrow\cos(\omega_\mathrm{m}t+\phim)$.
	
	\subsection{In-coupling photonic efficiency and optomechanical coupling}\label{sec:in_coup}
\begin{figure}[t]
		\centering
		\includegraphics[width=\columnwidth]{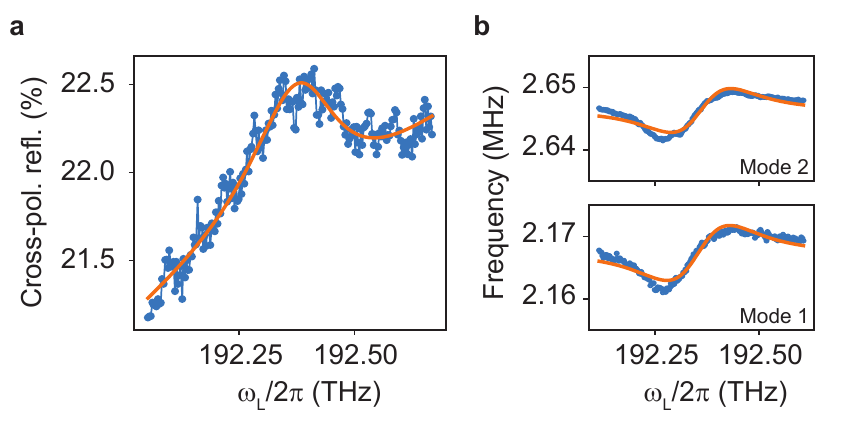}
		\caption{\textbf{Estimation of $\kin$ and $g_0$.} \textbf{a}, Measured reflection spectrum (blue points) along with the fitted response (orange curve) obtained as a function of the drive laser detuning in the absence of the detection laser. The in-coupling rate is obtained from the fit. $P_\mathrm{in}=\SI{126.7}{\micro\watt}$. \textbf{b}, The mechanical frequencies of the modes (blue data points) are obtained by a Lorentzian fits to the thermo-mechanical noise spectra (shown in the main manuscript). The optical spring shift of the two mechanical modes, as measured by the detection laser, is fitted (orange curve) to extract $g_{0i}$. $P_\mathrm{in}=\SI{33.4}{\micro\watt}$.}
		\label{SIfig1}
	\end{figure}

	The in-coupling rate to the cavity can be estimated from the reflection spectra obtained as the drive laser is detuned across the cavity resonance while keeping the laser power constant. Figure \ref{SIfig1}\textbf{a} shows the measured cross-polarized reflection from the cavity in the absence of the detection laser. The reflection spectrum shows the cavity resonance as a Fano lineshape which arises from an interference of the resonant and non-resonant contributions to the reflection. The overall slope in the background arises external to the device and is attributed to optical components in the setup. The cross-polarized reflectance is fitted to an equation of the form \cite{SIleijssen2015strong}
	\begin{equation}
	R = \left|c e^{i\theta}-\frac{\sqrt{\kin\kout}}{-i\Delta+\kk/2}\right|^2 + b\Delta,
	\end{equation} where $\kin$, $\kout$ are the in-, out-coupling rates of the cavity, $c$ and $\theta$ are the amplitude and phase of the non-resonant contribution to the reflection due to direct scattering, and $b$ is the slope of the background. For the cross-polarized measurement scheme used here, the in-coupling rate is bounded by $\kin\leq\sqrt{\kin\kout}$ \cite{SIleijssen2015strong}. The upper bound assumes that the in-/out-coupling rates of the cavity are equal, where the fitted response gives $\kin/\kk=0.416\%$.
	
	The presence of the drive laser modifies the mechanical response of the two modes according to the optical spring shift. The optical spring effect can be fitted to the equation
	\begin{equation}
	\delta\Omega_i = g_{0i}^2 \kin \frac{P_\mathrm{in}}{\hbar\ooL}\frac{2\Delta}{(\Delta^2+\kk^2/4)^2},
	\end{equation} where $g_{0i}$ are the single-photon optomechanical coupling strengths and $P_\mathrm{in}$ is the drive laser power. The mechanical frequencies of the two modes along with the fitted optical spring shift response is shown in \autoref{SIfig1}\textbf{b} with fitted parameters $g_{01}/2\pi=\SI{18.5}{\mega\hertz}$ and $g_{02}/2\pi=\SI{15.4}{\mega\hertz}$.

	\subsection{Phonon conversion - detuning dependence}
\begin{figure}[t]
		\centering
		\includegraphics[width=\columnwidth]{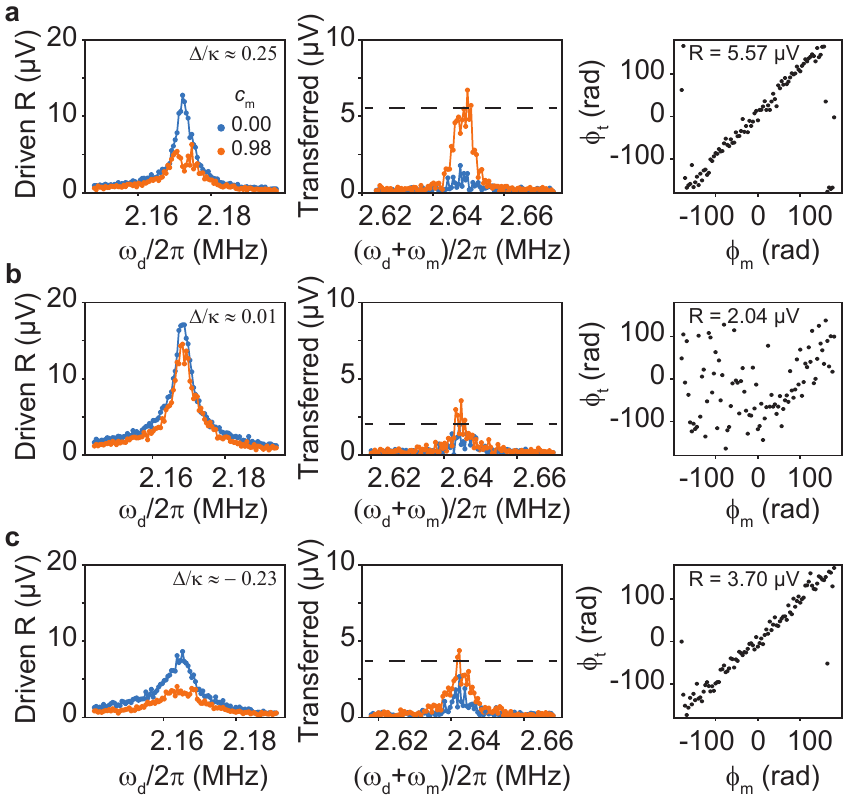}
		\caption{\textbf{Detuning dependent phonon transfer.} \textbf{a},\textbf{b},\textbf{c}, Mode transfer measurements for a detuning of $\Delta/\kk\simeq$ 0.25, 0.01, and -0.23 respectively. The left two panels show the driven response of mode 1 and simultaneously measured transferred response to mode 2 for a fixed $\om$ chosen at each detuning. The phase imprint on mode 2 as a function of the modulation phase is shown in the rightmost panels. There is no clear phase pickup for near zero detuning of the drive laser. The dashed line in the transferred response corresponds to the average magnitude of the transfer signal measured during the phase transfer measurement.}
		\label{SIfig2}
	\end{figure}

	The mode conversion experiments shown in the manuscript are performed at a drive laser detuning chosen such that the optical spring shift is large. The maximum optical spring shift, and hence the largest intermodal coupling strength, corresponds to a detuning of $\Delta/\kk=\pm1/2\sqrt{3}\simeq\pm0.289$. This is obtained from the functional form of $g_\mathrm{eff}$ incorporating the detuning dependence of the average photon population. It is also evident from the form of $g_\mathrm{eff}$ that the mode conversion strength is zero for zero detuning. This detuning dependence is observed in the plots shown in \autoref{SIfig2}. The driven response of mode 1 is largest when the drive laser is resonant with the cavity (\autoref{SIfig2}\textbf{b}). However, the transferred response is seen to be negligible compared to the transfer obtained for other detunings (\autoref{SIfig2}\textbf{a},\textbf{c}). This is also evident from the measurements showing the transferred phase where there is no clear phase pickup at mode 2 for zero detuning. The optical power of the drive laser is not kept constant for the different detunings used in \autoref{SIfig2}.
	
	%
	
\end{document}